\documentclass[11pt,a4paper]{amsart}
\usepackage{amssymb}
\usepackage{amsmath,amsthm}
\usepackage[toc,page]{appendix}
\usepackage[pagebackref]{hyperref} 
\usepackage{datetime}

\newtheorem{theorem}{Theorem}
\theoremstyle{definition}

\newcommand{\ii}{{\mathrm{i}}}
\newcommand{\R}{\mathbb{R}}

\newcommand{\U}{U_+}

\newcommand{\dd}{{\mathrm{d}}}

\newcommand{\bX}{{\bf X}}

\newcommand{\cU}{{\mathcal{U}}}
\newcommand{\cV}{{\mathcal{V}}}

\begin{document}

\title[Regularity of geodesics]{The regularity of geodesics in impulsive
\emph{pp}-waves}

\author[A. Lecke]{Alexander Lecke}
\address{Faculty of Mathematics, University of Vienna, Oskar-Morgenstern-Platz
1, 1090 Vienna, Austria}
\email{alexander.lecke@univie.ac.at}
\author[R. Steinbauer]{Roland Steinbauer}
\address{Faculty of Mathematics, University of Vienna, Oskar-Morgenstern-Platz
1, 1090 Vienna, Austria}
\email{roland.steinbauer@univie.ac.at}
\author[R. Svarc]{Robert \v{S}varc}
\address{Institute of Theoretical Physics, Charles University in Prague,
Faculty of Mathematics and Physics, V~Hole\v{s}ovi\v{c}k\'ach~2, 180~00 Praha
8, Czech
Republic }
\email{robert.svarc@mff.cuni.cz}

\begin{abstract}
We consider the geodesic equation in impulsive \emph{pp}-wave space-times
in Rosen form, where the metric is of Lipschitz regularity. We prove that
the geodesics (in the sense of Carath\'eodory) are actually continuously 
differentiable, thereby rigorously justifying the ${\mathcal C}^1$-matching
procedure which has been used in the literature to explicitly derive the
geodesics in space-times of this form.
\medskip 

\emph{PACS 2010:} 04.20.Jb, 
02.30.Hq 

\end{abstract}

\maketitle

\section{Introduction}

Impulsive \emph{pp}-waves (\cite{Penrose:72}) have become text-book examples of
exact solutions modeling gravitational wave pulses, see \cite[Ch.\ 20]{GP:09}
and \cite{P02} for an overview. They can be described by the line element in
Brinkmann form 
\begin{equation}\label{mbf}
\dd s^2=2H(\zeta,\bar \zeta)\delta(\cU)\dd \cU^2-2\dd \cU\dd \cV
+2\dd \zeta\dd\bar \zeta,
\end{equation}
where for convenience we have used complex coordinates 
\begin{align*}
\zeta=\frac{1}{\sqrt{2}}\big(x+\ii y\big) \,, \qquad\qquad
\bar{\zeta}=\frac{1}{\sqrt{2}}\big(x-\ii y\big) \,,
\end{align*} 
and $(\cU,\cV,x,y)\in\R^4$. Here $H$ is a real-valued function of the
spatial variables which we assume to be smooth (except for possible
singu\-la\-ri\-ties which we then remove from the space-time) 
and $\delta$ denotes the Dirac-function. 
In these coordinates the metric takes manifestly Minkowskian 
form in front and behind the wave impulse which is located on the 
null hypersurface $\{\cU=0\}$. This, however, comes at the expense of
introducing a \emph{distributional} coefficient into the metric. Alternatively
the space-time is described in Rosen form 
(\cite{Penrose:72,DEa78,PV98})
\begin{align}\label{eq:mrf}
   \dd s^2 = 2\left|\dd Z+\U\big(H_{,Z\bar{Z}}\dd Z+ H_{,\bar{Z}\bar{Z}}\dd
\bar{Z}\big)\right|^2-2\dd U\dd V,
\end{align}
where again we have used complex coordinates in the transverse space
\begin{align*}
Z=\frac{1}{\sqrt{2}}\big(X+\ii Y\big) \,, \qquad\qquad
\bar{Z}=\frac{1}{\sqrt{2}}\big(X-\ii Y\big) \,,
\end{align*} 
 and $(U,V,X,Y)=(U,V,X^2,X^3)\in\R^4$. Moreover,
\begin{align*}
\U(U) &= 
    \begin{cases}
     0 & \text{if } U \leq 0, \\
     U & \text{if } U \geq 0
    \end{cases}
\end{align*}
denotes the kink-function and hence the metric \eqref{eq:mrf} is \emph{Lipschitz
continuous}.

The geodesics of \eqref{mbf}, which actually are broken and 
refracted straight lines with a jump in the $\cV$-coordinate, 
have been derived in \cite{FPV88}, while in
\cite{B97,geo} the geodesic equations (which are non-linear 
ODEs with distributional right hand sides, hence mathematically delicate) 
have been treated rigorously. Finally in \cite{KS99} the geodesic
equations of \eqref{mbf} have been proven to possess unique global solutions
in a suitable space of (non-linear) generalized functions (\cite{GKOS01,C85}).  
This result in turn  enabled a mathematically sensible 
treatment (\cite{KS99a,EG11}) of the discontinuous ``coordinate transform``
introduced by Penrose (\cite{Penrose:72}) 
which relates \eqref{mbf} and \eqref{eq:mrf} (see also
\eqref{DiscontTransReal}, below).

On the other hand the geodesics in space-times similar to \eqref{eq:mrf}
(impulsive \emph{pp-}waves (\cite{PV99,bis-proc}), non-expanding (Kundt)
impulsive waves with a cosmological constant (\cite{PG99,PO01}), and expanding
impulsive waves (\cite{PS03,PS10})) 
have been derived by pasting together the geodesics of the background in a
${\mathcal C}^1$-manner. More precisely, in our case assuming the geodesics to
be ${\mathcal
C}^1$-curves in the continuous metric \eqref{eq:mrf} one may match the straight
line solutions given in manifestly Minkowskian coordinates on either side of
the wave to obtain explicit global geodesics (see also section
\ref{sec:matching}, below). While this ``${\mathcal C^1}$-matching procedure''
basically gives the correct answer (\cite[Sec.\ 4]{bis-proc}) the key assumption
that allows for the matching at all has remained unproven. In fact, the
Christoffel symbols of the Lipschitz continuous metric \eqref{eq:mrf} and hence
the right hand side of the geodesic equations are only locally bounded but
discontinuous, and at first sight the ${\mathcal C}^1$-property as well as
uniqueness of the geodesics seems to be too much to hope for. 
   
In this short note, we \emph{prove} that the geodesic equation of \eqref{eq:mrf}
actually possesses unique ${\mathcal C}^1$-solutions. To this end we employ the most natural
solution-concept available for ODEs with discontinuous right hand sides of
this form, which is due to Cara\-th\'{e}o\-dory (see e.g.\ \cite[Ch.\ 1]{F:88}).
It is a minimal extension of the classical solution concept and provides an
existence and uniqueness theorem for systems of the form
\[\dot x(t)=f(t,x(t))\]
basically assuming $f$ to be Lipschitz continuous
only with respect to $x$ and merely measurable w.r.t.\ $t$. 
Moreover the solutions are guaranteed to be absolutely continuous. For the 
convenience of the reader we have collected the basic facts on
Carath\'{e}odory solutions in an appendix, for all details we refer to the
literature. 

We prove the ${\mathcal C^1}$-property of the geodesics in section \ref{sec:sol}
and using the (now justified) matching procedure derive an explicit
description of the geodesics in generic impulsive \emph{pp}-waves in section
\ref{sec:matching}.

\section{The regularity of geodesics for impulsive pp-waves}\label{sec:sol}

In this section we explicitly calculate the geodesic equations for impulsive
\emph{pp}-waves in the continuous form of the metric \eqref{eq:mrf} and
demonstrate that the coefficients obey the assumptions of Carath\'{e}odory's 
existence and uniqueness theorem (Theorem \ref{thm:cara}). In this way we
prove that the (Carath\'{e}odory) solutions of the geodesic equations are
continuously differentiable.

We start by rewriting metric \eqref{eq:mrf} in real form 
\begin{align}\label{eq:metric_real}
 \dd s^2 = g_{ij}\,(U,X^k)\,\dd X^i \dd X^j -2\,\dd U \dd V,
\end{align}
where the spatial metric is given by
\begin{align}\label{eq:sm}g_{ij}=\delta_{ij}+2\U
H_{,ij}+(\U)^2\delta^{kl}H_{,ik}H_{,jl}
\end{align} 
for $i,j,k,l=2,3$. So the $g_{ij}$ and hence the entire metric is smooth
w.r.t.\ $X^i$ but merely Lipschitz continuous w.r.t.\ $U$.  
Recall that by Rademacher's theorem (locally) Lipschitz continuous functions 
 are differentiable almost everywhere with derivative belonging (locally) to $L^\infty$.
Taking derivatives of the metric coefficients will always be understood in this
sense. Therefore we deliberately do not denote the kink function by
$U\Theta(U)$ to avoid any confusion which might arise from using 
multiplication rules on $(U\Theta(U))^2$ (possibly obscuring the fact that
$U_+^2\in
{\mathcal C}^1$).

The non-vanishing Christoffel symbols are
\begin{equation}\label{eq:christ_non_van}
 \Gamma^V_{jk} = \frac{1}{2}g_{jk,U} \,, \qquad \Gamma^i_{Uk} =
 \frac{1}{2}g^{ij}g_{jk,U} \,, 
 \qquad \Gamma^i_{kl},
\end{equation}
where $\Gamma^i_{kl}$ $(i,k,l=2,3)$ are the Christoffel symbols of the spatial
metric \eqref{eq:sm}.
Since all the Christoffel symbols of the form $\Gamma^U_{\mu\nu}$ vanish we may
use $U$ as an affine parameter for the geodesics. Setting $\dot U=1$ 
the geodesic equations take the form

\begin{eqnarray}
\ddot{V} +\frac{1}{2}g_{ij,U}\dot{X}^i\dot{X}^j = 0 
\,, \quad \ddot{X}^i +\Gamma^i_{kl}\dot{X}^k\dot{X}^l
+g^{ij}g_{jk,U}\dot{X}^k = 0 
\end{eqnarray}
and after some calculations we explicitly obtain
\begin{eqnarray}\label{eq:geo-expl}
\ddot{V}&=&
-\big[(U_+)_{,U}\,H_{,ij}+\textstyle{\frac{1}{2}}(U_+^2)_{,U}\,\delta^{mn}H_{,im
}H_{,jn}\big]\dot{X}^i\dot{X}^j\,, \nonumber \\
\ddot{X}^i&=&
-g^{ij}\big[
U_+\,H_{,jkl}+U_+^2\,\delta^{mn}H_{,jm}H_{,kln}\big]\dot{X}^k\dot{X}^l 
\\
&&-2
g^{ij}\big[
(U_+)_{,U}\,H_{,jk}+\textstyle{\frac{1}{2}}(U_+^2)_{,U}\,\delta^{mn}H_{,jm}H_{,
kn}\big]\dot{X}^k \,,\nonumber
\end{eqnarray}
where the coefficients of the inverse spatial metric are of course given by
\begin{align}\nonumber
g^{ij}=D^{-1}g_{pq}(\delta^{ij}\delta^{pq}-\delta^{ip}\delta^{jq}\big),\ 
D\equiv\det g_{ij}
=
\frac{1}{2}\big(\delta^{ij}\delta^{pq}-\delta^{ip}\delta^{jq}\big)g_{ij}g_{pq}.
\end{align}

We now interpret system \eqref{eq:geo-expl} as a first order
system in the dependent variables $\bX:=(V, \tilde V=\dot V, X=X^2, \tilde
X=\dot
X^2, Y=X^3,\tilde Y=\dot X^3)$ and the independent variable $U$ and 
check that the conditions of Theorem \ref{thm:cara} (see appendix) 
are satisfied. Since these
are obviously fulfilled for the trivial equations $\dot V=\tilde V$, $\dot
X=\tilde X$ and $\dot Y=\tilde Y$ we are left with the task of
verifying conditions (A)--(C) of Theorem \ref{thm:cara} for the
right hand sides of \eqref{eq:geo-expl}, locally around every
point $p$ lying on the shock surface $\{U=0\}$. 

Clearly every such point $p$ has some neighborhood ${\mathcal W}$ where  $D$
is bounded away from $0$, hence the inverse spatial metric on ${\mathcal W}$ is
smooth w.r.t.\ $(X,Y)$ and Lipschitz continuous w.r.t.\ $U$. Consequently the
r.h.s.\ of \eqref{eq:geo-expl} is smooth w.r.t.\ $\bX$ and
(due to the terms $(U_+),_{U}=\Theta(U)$ merely) $L^\infty$ w.r.t.\ $U$ on
${\mathcal W}$. This, however, gives (A) (with the exceptional value $U=0$),
(B),  and (C) (with ${\mathfrak m}$ actually in $L^\infty$) on
${\mathcal W}$.

Now given arbitrary data at $p$, Theorem
\ref{thm:cara} provides us with a unique solution $\bX$ locally around the
shock hypersurface. In addition $\bX$ is absolutely continuous which implies
that the velocities $(\dot V,\dot X,\dot Y)$ are continuous. Hence the
geodesics are ${\mathcal C}^1$. Since off the shock hypersurface the space-time
is just Minkowski space we may match the solutions obtained above to the
geodesics of the background to obtain global solutions.

Hence we have shown that impulsive \emph{pp}-waves in the continuous form
possess unique global ${\mathcal C}^1$-geodesics. More precisely we may state
the following theorem.

\begin{theorem}\label{thm:thm}\
The geodesic equations for the impulsive \emph{pp}-wave metric \eqref{eq:mrf}
are uniquely globally solvable in the sense of Carath\'{e}odory and the
solutions are continuously differentiable. (In fact they possess absolutely
continuous velocities).
\end{theorem}

Finally we remark that our method crucially depends on the
fact that the coordinate $U$ can be used as an affine parameter for the
geodesics. This is, however, not the case for more general classes of impulsive
gravitational waves such as non-expanding impulsive waves on (anti) de-Sitter
background (\cite{PG99,PO01}) 
as well as expanding impulsive waves in all constant curvature
backgrounds (\cite{Penrose:72,H90,NP92,H92,H93,PG99a,PG00,AN01}). In this
situation the
geodesic equations have to be treated as an autonomous system of ODEs and in
this case Carath\'{e}odory's theorem provides no advantage over the classical
theory, i.e., it also needs the right hand side to be Lipschitz continuous; but
of course the Christoffel symbols will again only be bounded. A thorough
investigation of this case is subject to current research.

\section{${\mathcal C}^1$-matching}\label{sec:matching}

Using Theorem \ref{thm:thm} we now apply the ${\mathcal C}^1$-matching
procedure outlined in the introduction to derive the explicit
form of the geodesics for impulsive \emph{pp}-waves.
We start with Minkowski space-time in the form 
\begin{align}\label{ConfFlatCoordReal}
\dd{s}^2 = \dd x^2+\dd y^2 -2\dd\cU\dd\cV
\end{align}
and consider the (formal) transformation (\cite[eq.\ (20.4)]{GP:09})
\begin{align}\label{DiscontTransReal} \nonumber
 \cU &= U,\\
 \cV &=V+\Theta(U)H
 +\textstyle{\frac{1}{2}}\U\big((H_{,X})^2+(H_{,Y})^2\big),\nonumber\\
 x &= X+\U H_{,X},\\ \nonumber
 y &= Y+\U H_{,Y},
\end{align}
which is discontinuous in $\cV$ at $U=0$ and exactly gives Penrose's junction
conditions used in his ``scissors and paste'' approach (\cite{Penrose:72}). 
If we employ the transformation separately in the regions $U<0$ and $U>0$ we
obtain the continuous line element \eqref{eq:metric_real}. Observe that if one
formally
applies \eqref{DiscontTransReal} for all $U$ one obtains the distributional form
\eqref{mbf} of the metric (a procedure which has been made mathematically
precise in \cite{KS99a,EG11}).

We now stay with the continuous form \eqref{eq:mrf} of the metric and
apply Theo\-rem \ref{thm:thm} to obtain global $\mathcal{C}^1$-geodesics which we
denote by 
\begin{align}
 V=V(U) \,, \qquad X=X(U) \,, \quad Y=Y(U)\,,
\label{GeodesicsReal}
\end{align}
again using $U$ as an affine parameter. 
We now employ transformation  \eqref{DiscontTransReal}
separately for ${U<0}$ and ${U>0}$ and consider the geodesics
\eqref{GeodesicsReal} in the manifestly Minkowskian ``halves'' on either side of 
the impulse. We denote their limits and
the limits of their velocities as we approach the impulse from the region
$U<0$ by
$$\cV_\mathfrak{i}^-,\ \dot\cV_\mathfrak{i}^-,\ x_\mathfrak{i}^-,\ \dot
x_\mathfrak{i}^-,\ y_\mathfrak{i}^-,\ \dot y_\mathfrak{i}^-$$
and the limits as we approach the  impulse from the region $U>0$ by
$$\cV_\mathfrak{i}^+,\ \dot\cV_\mathfrak{i}^+,\ x_\mathfrak{i}^+,\ \dot
x_\mathfrak{i}^+,\ y_\mathfrak{i}^+,\ \dot y_\mathfrak{i}^+.$$ 
Here the subscript $\mathfrak{i}$ stands for ``(time of) interaction''.
Now the ${\mathcal C}^1$-property of the geodesics \eqref{GeodesicsReal}
allows us to relate these sets of ``interaction parameters'' to one another.
From \eqref{DiscontTransReal} we explicitly obtain
\begin{align*}
\cV_\mathfrak{i}^- &= \cV_\mathfrak{i}^+-H_\mathfrak{i},\\
\dot{\cV}^-_\mathfrak{i}&=\dot{\cV}^+_\mathfrak{i}-H_{\mathfrak{i},X}\dot{x}
_\mathfrak{i}^+-H_{\mathfrak{i},Y} \dot{y} _\mathfrak{i}^++\textstyle{
\frac{1}{2}}\big((H_{\mathfrak{i},X})^2+(H_{\mathfrak{i},Y}
)^2\big), \\
x_\mathfrak{i}^- &= x_\mathfrak{i}^+,\\
\dot{x}^-_\mathfrak{i}&=\dot{x}^+_\mathfrak{i}-H_{\mathfrak{i},X},\\
y_\mathfrak{i}^- &= y_\mathfrak{i}^+,\\
\dot{y}^-_\mathfrak{i}&=\dot{y}^+_\mathfrak{i}-H_{\mathfrak{i},Y},
\end{align*}
where $H_\mathfrak{i}$, $H_{\mathfrak{i},X}$ and $H_{\mathfrak{i},Y}$ denote the
value of $H$ respectively of
its derivatives on the respective geodesic \eqref{GeodesicsReal} at interaction
time $U=0$. So the
geodesics, \emph{as seen in the Minkowskian ``halves'' in front and behind the
wave}, suffer a jump in the $\cV$-component and are refracted in the
$\cV$-direction as well as in both spatial directions. 

These formulae coincide with the (distributional
limits of the) geodesics derived in the distributional form $\eqref{mbf}$ of
the metric in \cite[Thm.\ 3]{KS99} and we have thus given a second rigorous 
way of explicitly deriving the geodesics for impulsive \emph{pp}-waves.

\section*{acknowledgement}
We thank Ji{\v{r}}{\'{\i}} Podolsk\'y for kindly sharing his expertise and Clemens
S\"amann, and Milena Stojkovi\'c for helpful discussions. 
This work was supported by FWF-grant P25326 and OeAD WTZ-project CZ15/2013
resp.\ 7AMB13AT003.

\begin{appendix}
\section{Carath\'{e}odory solutions}
\renewcommand{\labelenumi}{(\Alph{enumi})}
\setcounter{theorem}{0}
\renewcommand{\thetheorem}{\Alph{section}\arabic{theorem}}

In this appendix we briefly summarize the basic facts of Carath\'{e}odory's
extension of the classical existence theory for ODEs and explicitly state the
theorem used to prove our main result in section \ref{sec:sol}. For all
details we refer to \cite[Ch.\ 1]{F:88} and \cite[Ch.\ 3 \S10, Suppl.\ 2]{W98}.

We consider the initial value problem for a non-autonomous system of 
first order ODEs 
\begin{align}\label{eq:ode}
 \dot{x}(t)=f(t,x(t)),\quad x(t_0)=x_0. 
\end{align}
Here $f:I\times D\to\R^d$, $I$ is an open interval containing $t_0$ and
$D\subseteq\R^d$ is an open and connected set which contains $x_0$.  

A \emph{Carath\'{e}odory solution} of \eqref{eq:ode} on an interval $J$ with
$t_0\in J\subseteq I$ is an absolutely continuous function
$x\colon J\to D$ which solves equation \eqref{eq:ode} almost
everywhere (in the sense of the Lebesgue measure) and
$x(t_0)=x_0$. We
recall that \emph{absolute continuity} is a strengthening of
continuity, which is weaker than Lipschitz continuity. More precisely,
a function $x\colon J\to \R^d$ is called absolutely continuous if for every
$\varepsilon>0$ there exists $\delta>0$ such that for every finite
sequence of pairwise disjoint sub-intervals $(a_k,b_k)$ of $J$ with
total length $\sum_k| b_k-a_k|<\delta$ we
have $\sum_k|x(b_k)-x(a_k)|<\varepsilon$.
Equivalently the derivative $\dot{x}$ of $x$ exists almost everywhere and we
have
\begin{align*}
 x(t)=x(t_0)+\int_{t_0}^t\dot{x}(\tau)\, \dd\tau.
\end{align*} 
Hence $x$ is a Carath\'{e}odory solution of \eqref{eq:ode} iff it
solves the equivalent integral equation
\[
 x(t)=x(t_0)+\int_{t_0}^tf(s,(x(s))\, \dd s.
\]

Of course every classical solution of \eqref{eq:ode} is a Carath\'{e}odory
solution
but the existence of the latter is guaranteed even for certain discontinuous
right hand sides $f$. More precisely, we have the following basic existence and
uniqueness theorem (cf.\ e.g.\ \cite[\S10, XVIII]{W98}).

\begin{theorem}\label{thm:cara} 
Let the function $f\colon I\times D\to\R^d$ satisfy the conditions
  \begin{enumerate}
   \item $f(t,x)$ is continuous in $x$ for almost all $t$,
   \item $f(t,x)$ is measurable in $t$ for all $x$,
   \item There exists $\mathfrak{m}\in L^1(I)$ with $|f(t,0)|\leq
     \mathfrak{m}(t)$ and 
    \[ |f(t,x)-f(t,y)|\leq \mathfrak{m}(t)\, |x-y|.
    \]
  \end{enumerate}
  Then there exists a unique (absolutely continuous) Carath\'eodory solution
  $x$ of \eqref{eq:ode} on some interval $J$ with $t_0\in J\subseteq I$.
\end{theorem}

\end{appendix}

\end{document}